\documentclass[conference]{IEEEtran}
\ifCLASSINFOpdf
\else
\fi
\usepackage{graphicx}
\usepackage{subfigure}
\usepackage{caption}
\usepackage{algorithm}
\usepackage{algorithmicx}
\usepackage{algpseudocode}
\usepackage{amsmath}
\usepackage{url}
\usepackage{flushend}

\floatname{algorithm}{Algorithm}

\begin{document}
\title{DeepProf: Performance Analysis for Deep Learning Applications via Mining GPU Execution Patterns}

\author{\IEEEauthorblockN{Jiazhen Gu\IEEEauthorrefmark{1},
Huan liu\IEEEauthorrefmark{1},
Yangfan Zou\IEEEauthorrefmark{1} and
Xin Wang\IEEEauthorrefmark{1}}
\IEEEauthorblockA{\IEEEauthorrefmark{1}School of Computer Science\\
Fudan University, Shanghai, China\\ Email:\{16210240029, 1521024081, zyf, xinw\}@fudan.edu.cn}}


%Telephone: (800) 555--1212, Fax: (888) 555--1212}


\maketitle

\begin{abstract}
Deep learning applications are computation-intensive and often employ GPU as the underlying computing devices. Deep learning frameworks provide powerful programming interfaces, but the gap between source codes and practical GPU operations make it difficult to analyze the performance of deep learning applications. In this paper, through examing the features of GPU traces and deep learning applications, we use the suffix tree structure to extract the repeated patten in GPU traces. Performance analysis graphs can be generated from the preprocessed GPU traces. We further present \texttt{DeepProf}, a novel tool to automatically process GPU traces and generate performance analysis reports for deep learning applications. Empirical study verifies the effectiveness of \texttt{DeepProf} in performance analysis and diagnosis. We also find out some interesting properties of Tensorflow, which can be used to guide the deep learning system setup.
\end{abstract}


% \fi
\IEEEpeerreviewmaketitle

\section{Introduction}
Recently machine learning techniques, especially deep learning \cite{deep}, have been proven effective to many sophisticated applications like audio recognition \cite{audio}, natural language processing \cite{nlp} and computer vision \cite{cv}. Many handy software frameworks, {\em e.g.}, Tensorflow \cite{tf}, have been proposed to improve the development efficiency of deep learning algorithms. With such tools, deep learning applications can be implemented with quite simple and concise codes. For example, with Tensorflow, a neural network to perform high-performance image recognition task can be realized in less than 100 lines of codes. Short as they are, such programs still require a long execution time. They are computation-intensive in nature, which are generally designed to process tremendous input data, and may incur lots of iterative runs. Bad design can result in intolerable execution time, which may even lead to their failed application in real-world scenarios. Therefore, execution efficiency is a critical concern of such programs.

However, analyzing the performance of deep learning applications is difficult. A key reason is that concise codes do not indicate the simplicity of the underlying computation. Actually, to get the best computational performance, deep learning applications usually employ NVIDIA GPUs (Graphics Processing Unit) as underlying computing devices \cite{nvidia}. Deep learning frameworks, like Tensorflow, provide simple upper-layer APIs (Application Program Interface) that wrap a sequence of complicated underlying GPU operations. When executing a deep learning application, the source codes will be optimized and transformed into concurrent GPU operations \cite{tfurl}. This may lead to a great gap between the codes written by developers and the operations actually executed. Therefore, the underlying execution details of applications are concealed from developers as well as potential performance issues, as a price for the convenience of programming.

Such a complex framework realization, along with the massive data transfer between host and GPUs, can be too complicated to comprehend even for an experienced machine learning developer. Moreover, current performance analysis tools are generally designed for applications running on CPU and cannot be used for GPU-based deep learning applications because the execution mechanism of GPU is different from that of CPU \cite{nvidia}. For example, the calls to CPU executions are blocked while computing operations are asynchronous on GPU, thus, CPU profiling cannot reflect the real execution procedure of GPU operations, not to mention performance analysis. On the other hand, Nvidia provides tools for GPU traces profiling \cite{nvprofiler}, but the results can hardly be understood by deep learning developers, who are typically unfamiliar with GPU operations. Therefore, how to bridge the execution gap between the program codes and the underlying execution, and to analyze the performance of deep learning applications is yet to be well-addressed.

In this paper, we first analyze the features of Tensorflow programs and the corresponding GPU traces. Based on the features, we summarize some execution rules of Tensorflow applications. We then use the rules to bridge the execution gap through recognizing repetition patterns in GPU traces. Since the size of GPU traces is generally large, we leverage suffix-tree structure \cite{weiner1973linear} to partition the unordered GPU operations in $O(n)$ space. By doing this, we simplify the tedious GPU traces into the granularity of iteration, since deep learning applications are typically iterative programs. Based on the second generated GPU profiles, we generate performance analysis graphs to help finding potential performance issues.

We implement \texttt{DeepProf} (Deep learning Profiler) to automatically process GPU traces and generate performance analysis of deep learning applications. Through \texttt{DeepProf}, we further summarize several execution properties of Tensorflow applications. In particular, we profile different Tensorflow applications under different GPU devices with \texttt{DeepProf}. The results reveal some potential GPU usage properties of Tensorflow applications, which can be used as guidance for system setup.

The rest of this paper is organized as follows. We introduce some preliminary knowledge on machine learning and deep learning frameworks in Section \uppercase\expandafter{\romannumeral2}, using Tensorflow as a typical example. Section \uppercase\expandafter{\romannumeral3} illustrates a motivating example. In Section \uppercase\expandafter{\romannumeral4}, we analyze the execution details of Tensorflow applications and elaborate how to extract patterns from GPU traces. We also describe the implementation of \texttt{DeepProf} in this section. Case studies are discussed in Section \uppercase\expandafter{\romannumeral5}. Section \uppercase\expandafter{\romannumeral6} presents related work. Finally, we conclude the paper in Section \uppercase\expandafter{\romannumeral7}.

\section{background knowledge}
In this section, we describe some preliminary knowledge on machine learning applications. First, we introduce the main procedure of machine learning applications. Then we illustrate how to use Tensorflow, a typical deep learning framework, to execute the program on GPUs.

\subsection{Machine Learning Applications}
Machine learning is a field that focuses on giving "computers the ability to learn without being explicitly programmed." \cite{alpaydin2014introduction} Machine learning applications generally construct algorithms that can learn from and make predictions on data \cite{machinelearning}. Through building a model from sample inputs, the algorithms are able to overcome the strictly static program instructions and make data-driven predictions or decisions \cite{pr}. Machine learning models are defined by many model parameters and to get the right values for all the parameters, iterative algorithms are essential. The iterative aspect of machine learning makes the models able to independently adapt when exposed to new data \cite{pr}. Consequently, a machine learning application must have at least one loop to processing input data and update models, which we call the \textit{training step}.

Deep learning is a class of machine learning algorithms which learn multiple levels of representation and abstraction that help to make sense of data \cite{deep}. Since deep learning applications are designed to process large amount of data, the data is partitioned into batches and execute batch by batch in the \textit{training step}. Thus iteration also plays an important role in deep learning applications.

\subsection{Tensorflow Applications}
A powerful framework is essential to developing deep learning applications. Tensorflow is one of the most popular deep learning frameworks \cite{tfurl}. With Tensorflow, developers and researchers can focus on designing deep learning algorithms, with no regards for the complex underlying operations. To achieve this, Tensorflow leverages data flow graphs to represent the computation procedure, as well as other deep learning frameworks \cite{theano}. A dataflow graph is consisted of nodes and edges. Nodes in the graph represent mathematical operators, while edges represent data arrays communicated between nodes. To execute a defined dataflow graph, Tensorflow first generates GPU computing operations according to the optimized dataflow graph, and then deploys the underlying program to GPU.

In general, Tensorflow applications can be abstracted into two parts, \textit{graph construction} and \textit{iterative training}. Tensorflow provides developers with powerful APIs to create dataflow graphs. In \textit{graph construction} part, developers construct the dataflow graph using Tensorflow APIs. The graph consists not only essential computing operations of the \textit{training step}, but also calculations of intermediate results or performance measurements. In \textit{iterative training} part, developers have to create a tensorflow \emph{session}, which contains the whole predefined dataflow graph. In order to execute the graph, calling \emph{session.run} method in a loop is required. The \emph{session} initializes the whole graph at the first time \emph{session.run} is called. Every \emph{session.run} call takes a list of parameters that assign the execution part in the dataflow graph. The session and dataflow graph mechanism allow developers to develop deep learning applications without any underlying details.

\subsection{CUDA}
As mentioned in previous sections, deep learning applications are actually executed on NVIDIA GPUs to get the most computation resources. The underlying computation platform used by most deep learning frameworks is CUDA \cite{cuda}. CUDA is a parallel computing platform provided by NVIDIA. Deep learning applications usually employ GPUs to accelerate execution time because the applications generally involve huge amount of matrix multiplications and other operations. These operations can be massively parallelized and thus sped up on GPUs. With CUDA, developers are capable of deploy their programs on NVIDIA GPUs for high performance computing. However, GPU based applications are much more complicated than normal CPU-based ones. To run programs on a GPU, developers have to deal with additional data management,({\em e.g.}, data copy from host to GPUs), as well as concurrency control. To realize a simple matrix multiplication requires more than 100 lines of C++ codes, not to consider sending massive data and executing complicated computations.

Tensorflow, like most other deep learning frameworks, hides complex underlying CUDA realization from developers. Every Tensorflow operators in the dataflow graph corresponds to one or more CUDA operations including computing and memory management. With Tensorflow, developers can focus on the design of deep learning algorithms and leave the intricate CUDA operations to the powerful framework. In next Section, we illustrate the urgency and difficulty of deep learning performance analyzers by a motivating example.

\section{a motivating example}
The dataflow graph mechanism of Tensorflow inevitably results in complicated underlying executions. Consequently, even simple codes may potentially contain subtle performance issues, which are particularly hard to be detected by current profiling tools. We demonstrate this by a simple example in Figure 1.

The program in Figure 1 intends to build a gradient descent model for recognizing images. The program creates the model through line 2 to line 6. After constructing the dataflow graph, a \emph{session} is required to launch the graph. In line 8, the program creates a Tensorflow session instance \emph{sess} and initializes all variables of the graph in line 9. Finally, the \textit{training step} of this program contains the defined \emph{train} and \emph{dbl\_loss} operations, by sending the operations as parameters to the \emph{sess.run} call. Since all the practical computations and underlying data copy are managed by the tensorflow session, we successfully implement the core algorithm within 20 lines of codes.

\begin{figure}[tb!]
\centering
\includegraphics[width=0.5\textwidth]{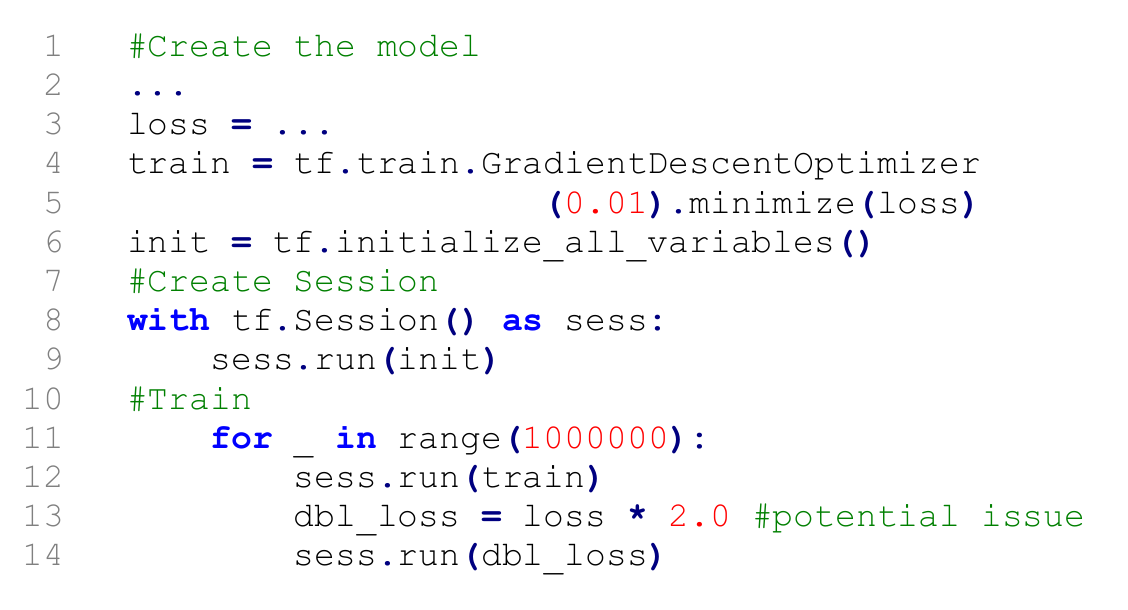}
\caption{An example of tensorflow program}
\label{fig_1}
\end{figure}

However, this simple program contains a subtle performance issue. In line 13, the overloaded operator ``*" implicitly calls \emph{tf.convert\_to\_tensor} function which leads to adding a new node to the graph. Since the \emph{sess} need to initialize the whole graph before computing, the program has to add a node and re-initialize the graph in every loop, which becomes the bottleneck of the performance. Without adding new nodes to the graph, the \emph{sess} only need to initialize the graph when the first time \emph{sess.run} is called. In the above example, the \emph{sess} has to initialize the graph at every loop step, which causes massive overhead to the execution time.

The performance issue is a typical memory leak problem caused by graph growth. The subtle operators that add nodes to the graph in training loop will result graph growth issues. It is however difficult to be unveiled. First, the issue is not easy to identify because the execution time depends on several factors. Underlying GPU performance and data transfer rate can affect the execution time as well. It is hard to tell whether the running time is abnormal or not because many developers cannot estimate the expected execution time precisely. Even if the performance issue is noticed, current tools cannot provide enough implications to help locate the problematic codes.

\textit{Nvprof}, provided by NVIDIA, is the official profiling tool to collect CUDA-related activities \cite{nvprof}. Although \textit{nvprof} is capable of listing all CUDA operations during the execution, the results are difficult for deep learning developers to understand. The GPU trace collected by \textit{nvprof} only contains GPU operations which have little relation to source codes of the program and make CPU profiling useless. Furthermore, the GPU traces generated by \textit{nvprof} can be quite large (58MB in our simple example). The human effort to analyzing the results manually is unaffordable for any individual or company.

Another tool provided by Tensorflow is \textit{timeline}. The \textit{timeline} tool is capable of profiling the Tensorflow operators in one \emph{sess.run} call and generate a .json file. However, the graph growth in the example happens before \emph{session.run} and hence the results are useless. Moreover, a simple application call \emph{sess.run} for thousands of times so that this tool will create thousands of json files. As a result, developers can't observe the overall performance of applications but the disconnected training step.

Therefore, lack of appropriate solutions to analyzing GPU traces makes it difficult to detect performance issues. In the next Section, through detailed analyzing Tensorflow applications and GPU traces, we propose \texttt{DeepProf}, a novel tool that helps developers to analyze deep learning application performance and to find potential problems.

\section{gpu trace analysis and processing}
To measure the application performance precisely, we expect to extract helpful information from the tedious and unreadable GPU traces. We first analyze the GPU traces in detail and extract the ordered GPU computing operations of Tensorflow applications. Leveraging the iterative property of deep learning algorithms, we use the suffix tree structure and an approximate matching algorithm to partition all computing operations to different parts. Every part refers to the underlying execution caused by one iteration step in deep learning algorithms. Based on the generated GPU profile with fine granularity, we define some metrics to measure the execution performance intuitively. We further implement \texttt{DeepProf} to automatically analyze the performance for deep learning applications. Although our approach can adapt to multi-loop scenarios, for simplicity, we only discuss the one-loop condition in analysis part. We show how to deal with multi-loop applications in the end of this section.

\newcommand{\tabincell}[2]{
\begin{tabular}{@{}#1@{}}#2\end{tabular}
}
\begin{table}[tb!]
  \begin{center}

   \caption{Attributes of GPU trace}\label{table1}
   \vspace{-1em}
        \begin{tabular}{ll}
            \\[-2mm]
            \hline\\[-2mm]
            {\bf \small Attribute}&\qquad {\bf\small Meaning}\\
            \hline\\
            \vspace{2mm}
            Start      &   \tabincell{l}{Time from the program starts to the operation is called}\\
            \vspace{2mm}
            Duration      &  \tabincell{l}{Execution time of the operation}\\
             \vspace{2mm}
            Size          &  \tabincell{l}{Data copy size }\\
             \vspace{2mm}
            Throughput  &   \tabincell{l}{Throughput of data copy operation}\\
             \vspace{2mm}
            Device    & \tabincell{l}{Device name the operation executed on}\\
            \vspace{2mm}
            Stream    & \tabincell{l}{Stream number the operation belongs to}\\
            \vspace{2mm}
            Name    & \tabincell{l}{Operation name}\\
            \hline
        \end{tabular}
        \vspace{-1.5em}
  \end{center}
\end{table}

\subsection{GPU Trace Analysis}
 Table 1 shows the attributes with their meanings of GPU traces collected through \textit{nvprof}. We introduce the concepts of CUDA operation and CUDA stream first. We classify the operations in GPU traces according to \textbf{name} and \textbf{stream} attributes and then summarize some important rules about the underlying execution of Tensorflow applications.

\subsubsection{CUDA Operations}
According to the \textbf{name} attribute, we notice that there are two main kinds of operations in GPU traces, \textit{kernels} and \textit{memcpy} (memory copy) operations. A \textit{kernel} is a function defined by CUDA api and can be executed on GPU. The \textit{kernels} can be defined by either CUDA or Tensorflow. The \textit{memcpy} operations can be further divided into three types according to the source and destination of the data copy: \textit{memcpyHtoD}, \textit{memcpyDtoH} and \textit{memcpyDtoD}. The 'H' means host while 'D' means GPU device. In the GPU trace of deep learning applications, we can directly distinguish two kinds of CUDA operations defined above, according to the \textbf{throughput} attribute. Specifically, \textit{computing} operations do not have \textbf{throughput} attribute while \textit{memcpy} operations do.

\subsubsection{CUDA Stream}
CUDA platform use stream mechanism to achieve concurrency. A \textit{stream} contains a sequence of operations that execute in issue-order on GPU \cite{cuda}. The operations in different streams may run concurrently and be interleaved. Note that although the operations in one stream are executed in order, one operation can utilize many processors and be executed in parallel on GPU. Figure 2 shows a simple concurrency example on GPU. After copying data to the GPU, the $Kernel$ and $MemcpyDtoH$ operations are executed by 4 streams and get 1.33x performance improvement than serial execution. Note that in $t$ moment, only $K1$ is executed on GPU and thus this operation can utilize all GPU processors for computing.

\begin{figure}[tb!]
\centering
\includegraphics[width=0.49\textwidth]{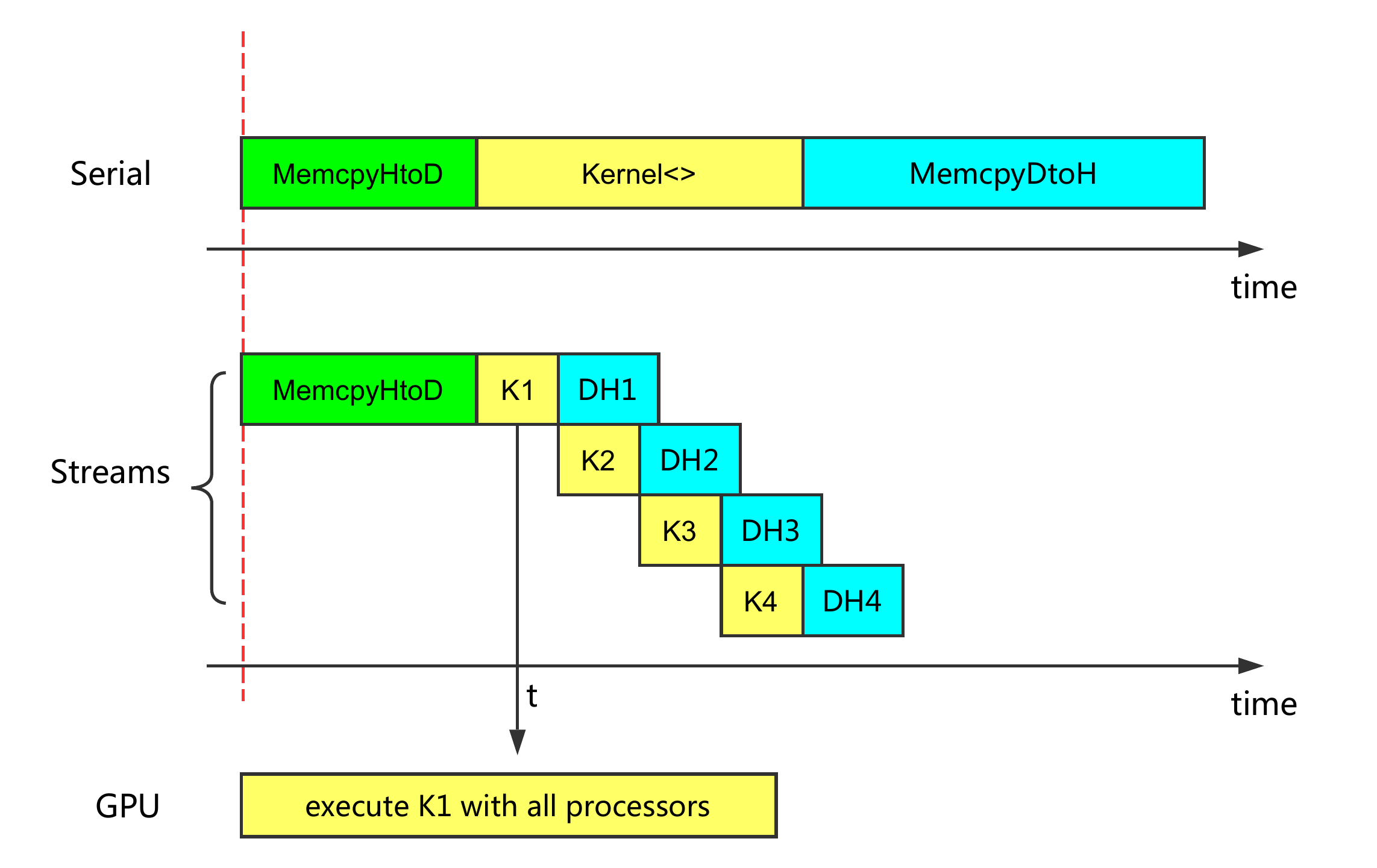}
\caption{An example on GPU concurrency}
\label{fig_2}
\end{figure}

Through the above analysis, we can reveal some underlying features of Tensorflow applications. As stated before, the operations in the same stream are executed in their issue-order. Through extracting all operations that have the same \textbf{stream} value and sorting them on the \textbf{start} value, we get the ordered operations of different streams during the execution. We continue to use the program in Section \uppercase\expandafter{\romannumeral3} as an example. After applying the extraction and sorting operations to the GPU trace, we summarize the distribution of different operations in each stream in figure 3. We notice that only stream 13 contains \textit{computing} operations while stream 14 and 15 only consists of \textit{memcpyHtoD} operations and \textit{memcpyDtoH} operations respectively. The rest stream (stream 7) has only two entries, one \textit{memset} (memory set) and one \textit{memcpyHtoD}. Based on above observation, we classify underlying streams used by TensorFlow applications into three kinds. The specific definitions are as follows:

1. \textit{Main stream}: Streams contain Tensorflow defined \textit{kernels}, like stream 13 in the example, are called main streams

2. \textit{Copy stream}: Streams only contain \textit{memcpy} operations, like stream 14 and 15, are called copy streams.

3. \textit{Assist stream}: All streams that do not belong to \textit{main stream} and \textit{copy stream} are called assist streams, like stream 7.

\begin{figure}[tb!]
\centering
\includegraphics[width=0.49\textwidth]{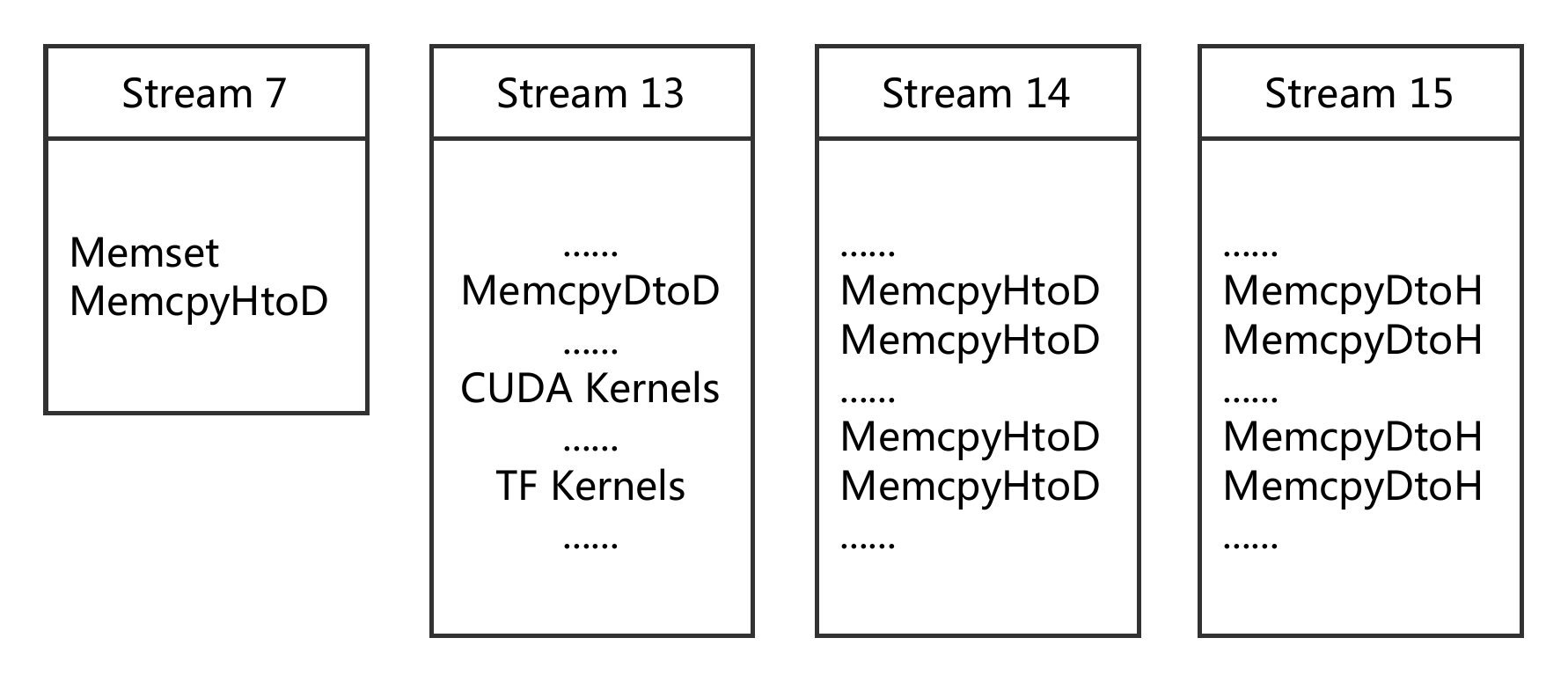}
\caption{An example of streams used by Tensorflow}
\label{fig_3}
\end{figure}

With above definitions on CUDA operations and streams, we further analyze the GPU traces of different Tensorflow applications, and summarize two general rules of the underlying Tensorflow realization as follows:

$Rule$ 1: Tensorflow only creates one \textit{main stream} \cite{tfurl}, which includes all \textit{kernels} defined by Tensorflow. Besides \textit{kernels}, this stream may contain \textit{memcpyDtoD} operations as well. This design allows the operations in \textit{main stream} to utilize all processers and memory spcace in GPU if necessary.

$Rule$ 2: Two \textit{copy streams} are created by Tensorflow to copy data between host and GPU. One stream deals with \textit{memcpyHtoD} operations while the other copys data from GPU to host.

$Rule$ 3: Assist streams are used to handle trivial GPU operations. The number of assist streams is depended on the realization of applications, but Tensorflow creates at least one assist stream, just like stream 7, to initialize GPU memory.

According to $Rule$ 1 and \textit{stream} mechanism of CUDA, we can conclude that the \textit{GPU computing procedure of Tensorflow is sequential}. Based on this conclusion, we leverage the iteration property of deep learning algorithms to partition the GPU traces.

\subsection{Suffix Tree Based Pattern Search}
Let's review the structure of Tensorflow applications. Any deep learning algorithm requires a \textit{training step}, which refers to a subgraph of the defined dataflow graph in Tensorflow. In \textit{iterative training} part, a session is created to execute the subgraph iteratively. We define the \textit{basic iteration} as one session.run on the subgraph of \textit{training step}. Once the dataflow graph is defined, the GPU operations caused by the \textit{basic iteration} is determined. According to $Rule$ 1, all the computing operations of Tensorflow are issued by one stream and thus have a strict execution order. As a result, executing \textit{basic iteration} will lead to the same underlying computing operations. Since applications execute the \textit{basic iteration} for many times, there must be a repeated pattern in the collected GPU traces. Note that the repeated pattern may not occur in every iteration because applications are likely to calculate some statistics periodically, which leads to the underlying operations different from those of \textit{basic iteration}. Under this circumstance, applications actually execute the \textit{basic iteration} with the operators performing statistic calculation. The GPU trace of such an iteration should be an insertion of some GPU operations to the \textit{basic iteration} trace.
Now we get the following criterion:
\newtheorem{theorem}{Criterion}
\begin{theorem}
The GPU trace consists of the basic iteration, which is interleaved by GPU operations for other purposes, such as statistic calculation.
\end{theorem}
Based on $Criterion$ 1, we can extract the \textit{basic iteration} caused operations through mining frequent patterns in GPU traces. We use the suffix tree structure to preprocess the original GPU traces.

\subsubsection{Suffix Tree}
Suffix trees allow efficient query processing on text data, which was first introduced in \cite{weiner1973linear}. Every internal node in a suffix tree represents a substring while every leaf node represents a suffix of the original string. Figure 4 shows a suffix tree of the string ``banana\$". The ``\$" symbol represents the string terminator. The example tree consists of 4 internal nodes and 7 leaves. Note that each node in the suffix tree represents a substring of the input string. More precisely, a leaf node corresponds to a suffix while a branch node corresponds to a prefix of the suffix, i.e., a substring. For example, the internal node ``5" in Figure 4 represents the substring "ana", and the leaf node ``9" represents the suffix ``anana". Furthermore every node and its descendants forms a subtree in the suffix tree. The number of leaves in a subtree determined by a node is the times that the corresponding substring repeats in the input string. For example, node ``5"  has 2 leaves in its subtree so ``Ana" repeats 2 times in ``banana".

\subsubsection{Why Suffix Tree}
As mentioned in section \uppercase\expandafter{\romannumeral3}, the GPU trace of a deep learning application is often extremely large. Both the iteration times and the complexity of algorithm can cause a great amount of underlying computations in GPU trace. Thus we cannot afford a high space complexity algorithm to discover frequent patterns. Constructing a suffix tree is quite efficient. The space and time complexity of constructing a suffix tree is $O(n)$, thus we choose suffix tree structure to deal with the massive original traces.

\begin{figure}[tb!]
\centering
\includegraphics[width=0.34\textwidth, height=0.22\textheight]{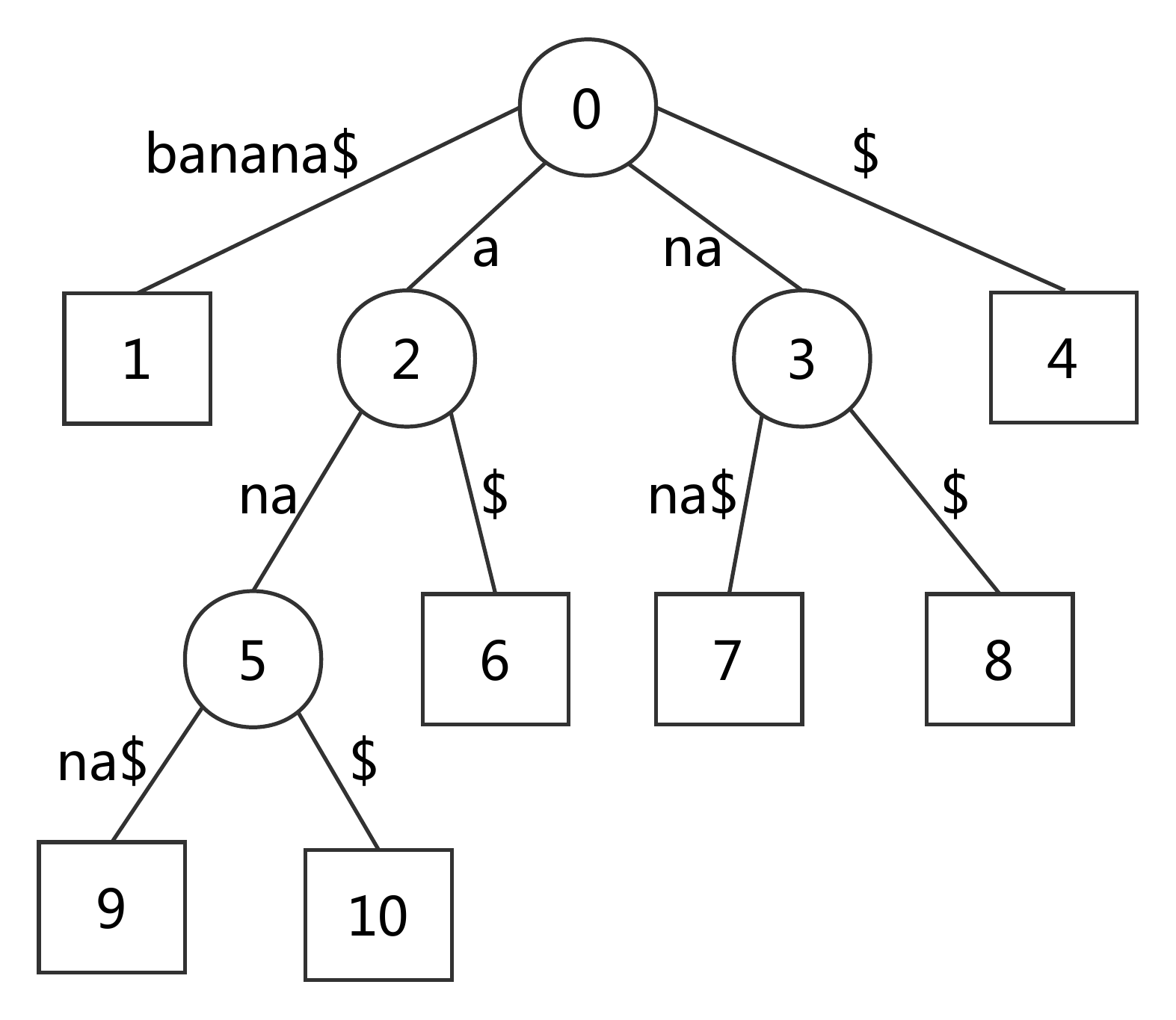}
\caption{The suffix tree of string ``banana\$".}
\label{fig_4}
\end{figure}

\begin{table}[tb!]
  \begin{center}

   \caption{Symbol Table}\label{table2}
   \vspace{-1em}
        \begin{tabular}{ll}
            \\[-2mm]
            \hline
            \hline\\[-2mm]
            {\bf \small Symbol}&\qquad {\bf\small Meaning}\\
            \hline
            \vspace{2mm}\\[-3mm]
            $S$      &   \tabincell{l}{The long string constructed by all operation names \\in a main stream}\\\\
            \vspace{2mm}
            $ST$          &  \tabincell{l}{ A suffix tree}\\
             \vspace{2mm}
            $l_{str}$          &  \tabincell{l}{The length of the string $str$}\\
             \vspace{2mm}
            $c_{str}$  &   \tabincell{l}{The repetition times of the string $str$ in $S$}\\
             \vspace{2mm}
            $l_p$    & \tabincell{l}{The length of the pattern}\\
            \vspace{2mm}
            $c_p$    & \tabincell{l}{The repetition times of the pattern}\\
            \vspace{2mm}
            $i$    & \tabincell{l}{The total number of iterations in a application}\\
            \vspace{2mm}
            $k_0$    & \tabincell{l}{The maximum unmatched string length}\\
            \hline
            \hline
        \end{tabular}
        \vspace{-1.5em}
  \end{center}
\end{table}

Next we introduce the details of pattern mining in GPU traces. For ease of reference, we summarize the symbols we use in Table 2.

We preprocess the original data and treat all operation names in the \textit{main stream} as a long string $S$. $S$ is a two-dimension array and every element of $S$ is a GPU operation name. We then construct a suffix tree $ST$ of $S$. Each substring repeats $c$ times in $S$ is represented by an internal node that has $c$ leaves as its descendants in $ST$. Moreover, the length $l$ of the repeated substrings should be appropriate because short patterns cannot form an iteration step while long patterns may contain too many operations. If we know the right $c_p$ and $l_p$ values, we can scan all nodes in $ST$ and select the frequent substring in $O(n)$ time.

\subsubsection{Parameter discussion}
The value of $c_p$ and $l_p$ are related to the number of \textit{basic iteration} executed. If the application executes exactly the same instructions in every iteration, $c_p$ is equal to $i$, which is the total number of iterations. If some iterative steps execute additional instructions, according to $Criterion$ 1, the number of total operations increases. Considering there are some initial operations, that do not belong to any iteration, the average GPU operations per iteration must be larger than $l_p$. Thus far we get the following two criterions for $c_p$ and $l_p$:
\begin{theorem}
$(i - \varepsilon) < c_p \leq i$, where $\varepsilon$ is the threshold we defined.
\end{theorem}
\begin{theorem}
$ l_p < \frac{l_S}{i}$, where $l_S$ is the total length of string S and $i$ is the number of iterations in the application. $\frac{l_S}{i}$ means the number of GPU operations per iteration.
\end{theorem}

Consequently, we can scan all nodes in ST and select frequent substrings that satisfy $Criterion 2$ and $3$. We only select the longest substring $P$ that repeats at least $i - \varepsilon$ times, and is shorter than $\frac{l_S}{i}$, where $\varepsilon$ is a priori threshold. If no internal node satisfies the filter conditions, the program doubles the value of $\varepsilon$ and re-search the suffix tree. In the worst scenario, the total time complexity of constructing ST and select the frequent substring is $O(mn)$, where $m$ is the re-search times for an eligible node in the tree.

Finally, we search for all patterns that approximately match $P$ in $O(mn)$ time, where m is the length of $P$ and n is the length of $S$, as shown in Algorithm 1. The parameter $k_0$ represents the maximum unmatched string length. In other words, the approximate patterns can have at most $k_0$ more operations than substring $P$. The algorithm checks all possible positions that may have a approximate match and return the start and end position of the matching substrings.
\makeatletter
\def\BState{\State\hskip-\ALG@thistlm}
\makeatother
\begin{algorithm}[tb!]
    \caption{Performance Metrics Table}
    \label{alg_2}
    \begin{algorithmic}[1]
        \Require substring $P$, string $S$
        \Ensure all matching positions in array $res$
        \State $res \gets \emptyset$
        \While{$i \leq length(S)$}
        \State $k \gets 0, i_0 \gets i$
            \For{ $j \gets 1 \cdots length(P)$}
                \While{ $S[i_0] \neq P[j]$ }
                    \State $k \gets k+1, i_0 \gets i_0+1$
                    \If{$k > k_0$} \State \textbf{goto} \emph{label} \EndIf
                \EndWhile
            \EndFor
        \BState \emph{label}:
            \If{$k \leq k_0$}
                \State $append(res, pair(i, i_0)); i \gets i_0+1$
            \Else
                \State $i \gets i+1$
            \EndIf
        \EndWhile
        \State \Return $res$
    \end{algorithmic}
\end{algorithm}

\subsection{Performance Metrics Analysis}
Through the suffix-tree-based approximate match algorithms, we successfully divide the ruleless GPU trace into different parts. Based on the partitioned trace, we further calculate a serious of metrics to measure the execution performance of deep learning applications.

We define the \textit{iteration interval} as the leisure time during two adjacent iterations. In other words, \textit{iteration interval} is the time difference between the start of one iteration and the end of its previous one. The latter iteration cannot start until GPU receives the input data and CPU finishes the other instructions in the loop of source codes, as the state graph shown in figure 5. Note that although data copy state is independent to GPU and CPU operations, the GPU iteration has to start after the corresponding data copy operations. Therefore, the size of \textit{iteration interval} represents the execution time of CPU instructions and data copy operations.

We further define the \textit{interval overlap} as the ratio of data copy time to \textit{compute interval}. We only consider the time of \textit{memcpyHtoD} operations during each \textit{iteration interval}, because copying data from GPU to host can be executed concurrently and has little effect on the start time of the next iteration. A high \textit{interval overlap} implies that the data transfer operations become the bottleneck. We also calculate the average value of \textit{iteration interval} and \textit{interval overlap}, as well as some intuitional metrics. Detailed meanings of different metrics are shown in Table 3.

\begin{figure}[htbp]
\centering
\includegraphics[height=0.25\textheight]{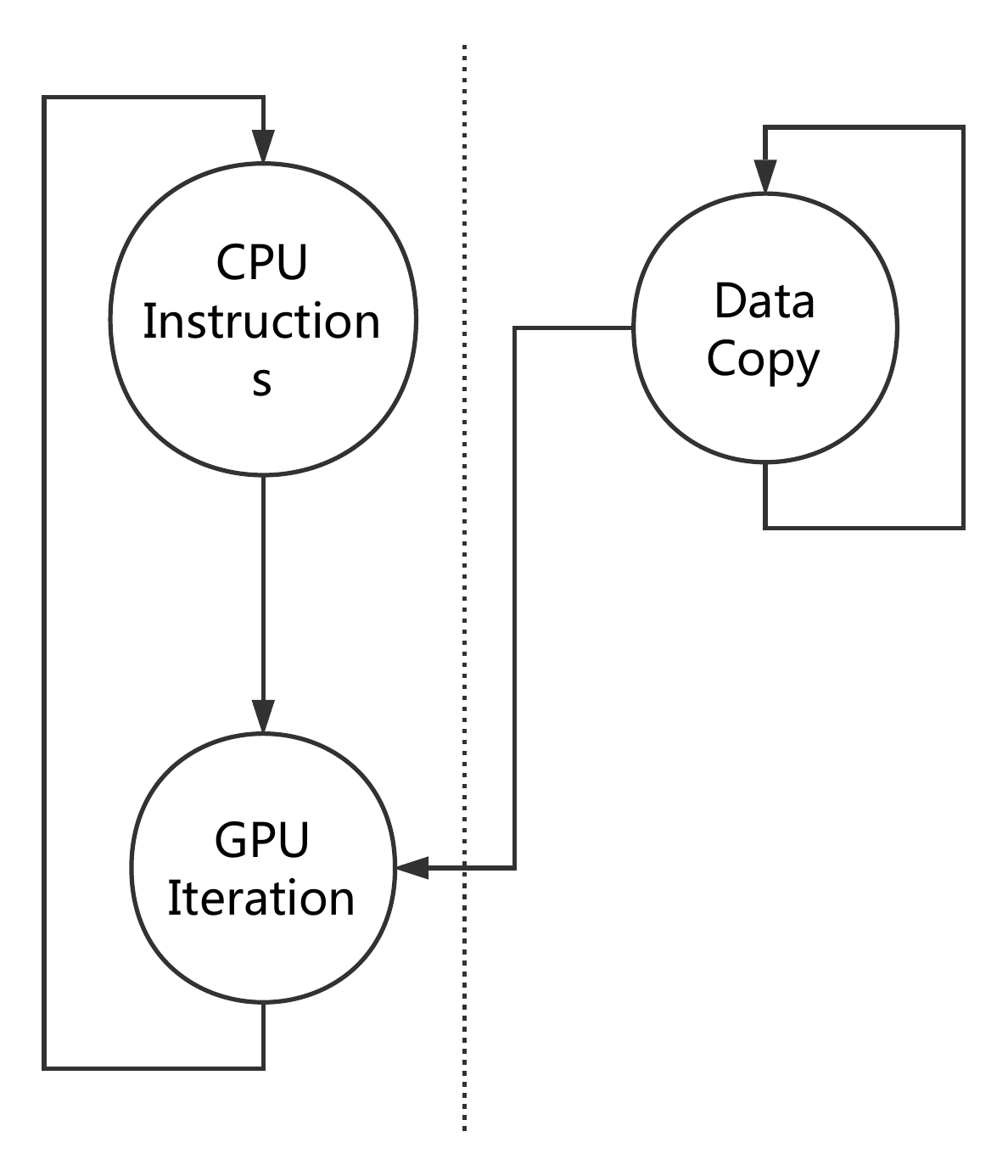}
\caption{State graph of iteration}
\label{fig_5}
\end{figure}

\begin{table}[tb!]
  \begin{center}

   \caption{Metrics Table}\label{table3}
   \vspace{-1em}
        \begin{tabular}{ll}
            \\[-2mm]
            \hline\\[-2mm]
            {\bf \small Metrics}&\qquad {\bf\small Meaning}\\
            \hline\\
            \vspace{2mm}
            avg interval   &   \tabincell{l}{Average of iteration interval}\\
            \vspace{2mm}
            max interval     &  \tabincell{l}{The maximum value of all iteration interval}\\
             \vspace{2mm}
            avg overlap     &  \tabincell{l}{Average value of interval overlap}\\
             \vspace{2mm}
            avg operation  &   \tabincell{l}{Average interval time of two GPU operations}\\
             \vspace{2mm}
            avg size    & \tabincell{l}{Average data copy size from host to GPU per iteration}\\
            \hline
        \end{tabular}
        \vspace{-1.5em}
  \end{center}
\end{table}

\subsection{\texttt{DeepProf} realization}
Based on the above discussion, we implement \texttt{DeepProf}, a novel tool to automatically mine patterns in original GPU traces and generate performance profiles for deep learning applications. The processing framework adopted in \texttt{DeepProf} is shown in figure 6. There are four main components of \texttt{DeepProf}. In preprocessing part, \texttt{DeepProf} first extract all GPU operations belong to the \textit{main stream} and then combine these operations into a long string $S$. In the pattern mining part, \texttt{DeepProf} construct the suffix tree of $S$, and mine the frequent pattern $P$ according to the number of iterations, which can be fetched through analyzing the source codes. The start and end positions of all approximately matched patterns of $P$ are generated in the approximate match part. According to the results from approximate match part, \texttt{DeepProf} partitions the GPU traces and calculates the performance metrics in the metrics generation part. The results generated by \texttt{DeepProf} contain a summary of average performance metrics and a file contains detailed metrics of every iteration. The results can be explained intuitively by the state graph shown in figure 6.

\begin{figure}[hb!]
\centering
\includegraphics[width=0.5\textwidth]{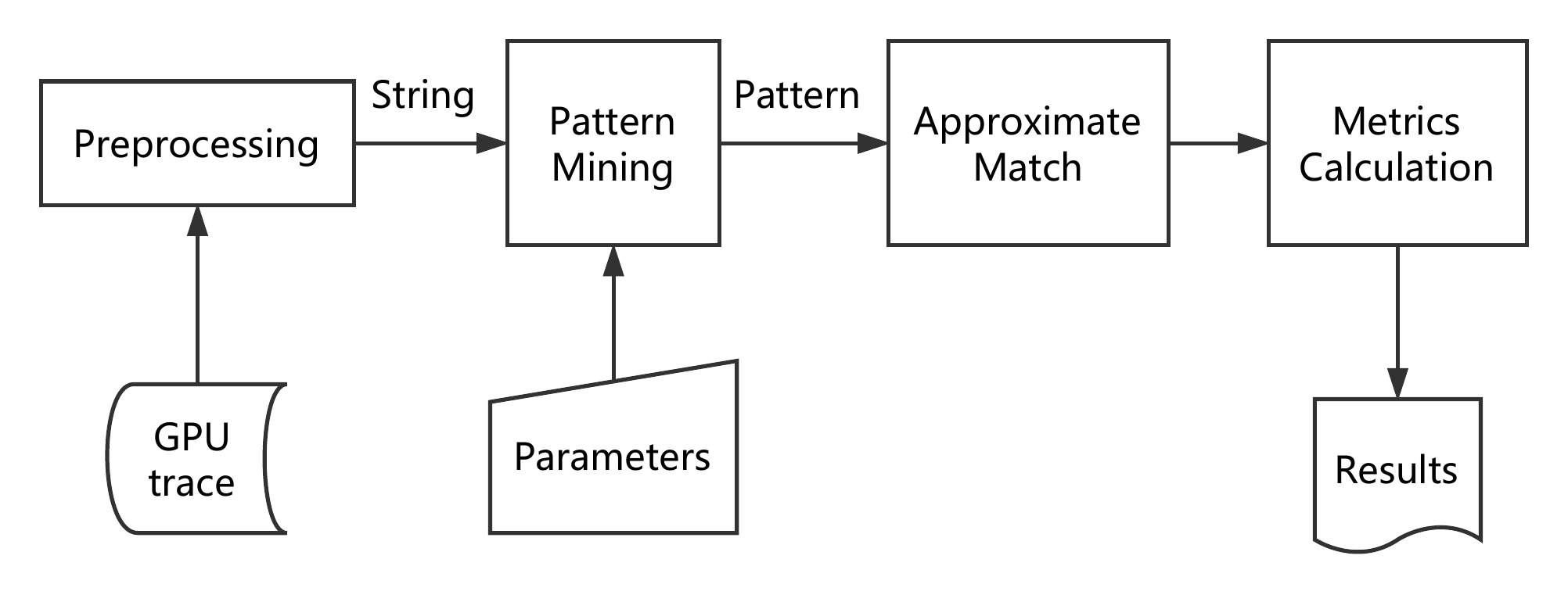}
\caption{Overview of \texttt{DeepProf} framework}
\label{fig_6}
\end{figure}

\subsubsection{Adapt to multi-loop}
The above \texttt{DeepProf} framework can adapt to more complicated applications with multiple loops, which means GPU traces can have several patterns. In the suffix tree, the multi-loop is property reflected in several frequent substrings. The repetition times and length of each substring still satisfy $Criterion 2$ and $3$. Knowing the number of loops and corresponding iterative steps, we just need to search the suffix tree with different parameters. All the parameters required to mine the frequent patterns can be obtained from the source codes, so \texttt{DeepProf} is capable of dealing with multi-loop applications.

Next, we use \texttt{DeepProf} to analyze and diagnose the performance of Tensorflow applications. The results show the effectiveness of \texttt{DeepProf}, and reveal some hidden features of Tensorflow.

\section{empirical study}
In this section, we evaluate the effectiveness of \texttt{DeepProf} in two aspects, performance analysis and diagnosis. We first analyze the performance of computation-intensive Tensorflow applications as well as IO-intensive ones on different GPUs by \texttt{DeepProf}. The results reveal a few execution features of Tensorflow, which can be used as guidance for system setup. The analysis results generated by \texttt{DeepProf} can also be used to diagnose performance issues. We demonstrate how to detect the potential performance issues according to the abnormal metrics using two cases. The first issue is shown in Section \uppercase\expandafter{\romannumeral3}, which is caused by graph growth. The second issue resides in data transfer procedure where oversize data copy becomes the bottleneck. We conduct the experiment on a Ubuntu 16.04 server with i7 6700K CPU and NVIDIA 1080Ti graphics card.

\subsection{Performance Analysis}
Tensorflow applications can be classified into two types, computation-intensive and IO-intensive. Computation-intensive applications employ a complex model to improve the prediction accuracy. In this kind of application, every training step requires a long execution time and the data transfer operations has little effect on performance. On the other hand, data-intensive applications are designed to process a large amount of input data, while large data transfer between host and GPU is common in this kind of applications. In general, developers intend to allocate the whole GPU memory to one application. In other words, one GPU only executes one application at a time. So our question is whether running several applications concurrently on one GPU will cause great performance loss. To address the problem, we first execute one application with a limited GPU memory size and use \texttt{DeepProf} to analyze the performance.

\begin{figure}[tb!]
  \centering
  \subfigure[Execution time with different memory size]{
    \label{fig:subfig:a} %% label for first subfigure
    \includegraphics[width=0.35\textwidth]{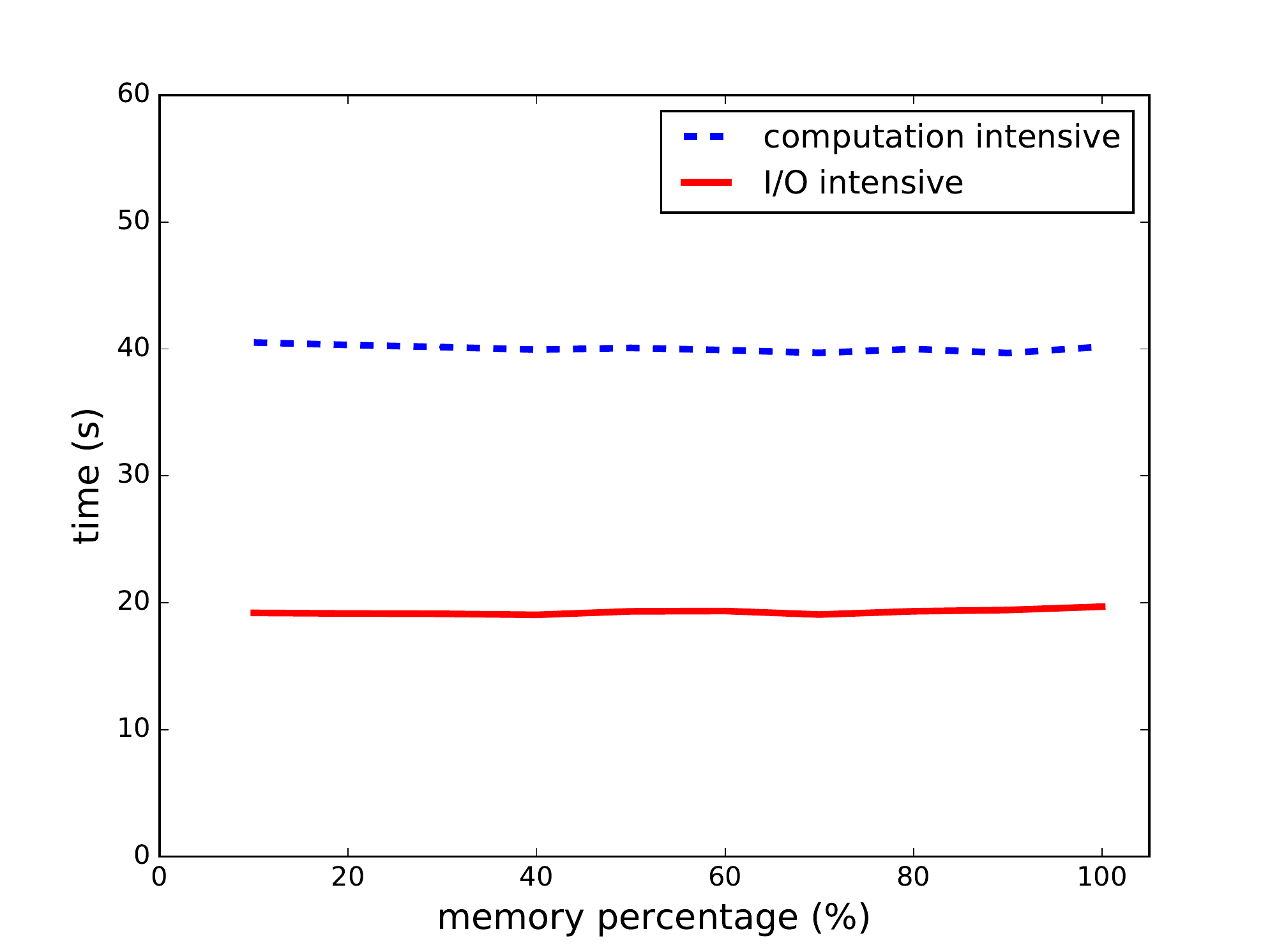}}
  \hspace{0.01in}
  \subfigure[GPU utilization with different memory size]{
    \label{fig:subfig:b} %% label for second subfigure
    \includegraphics[width=0.35\textwidth]{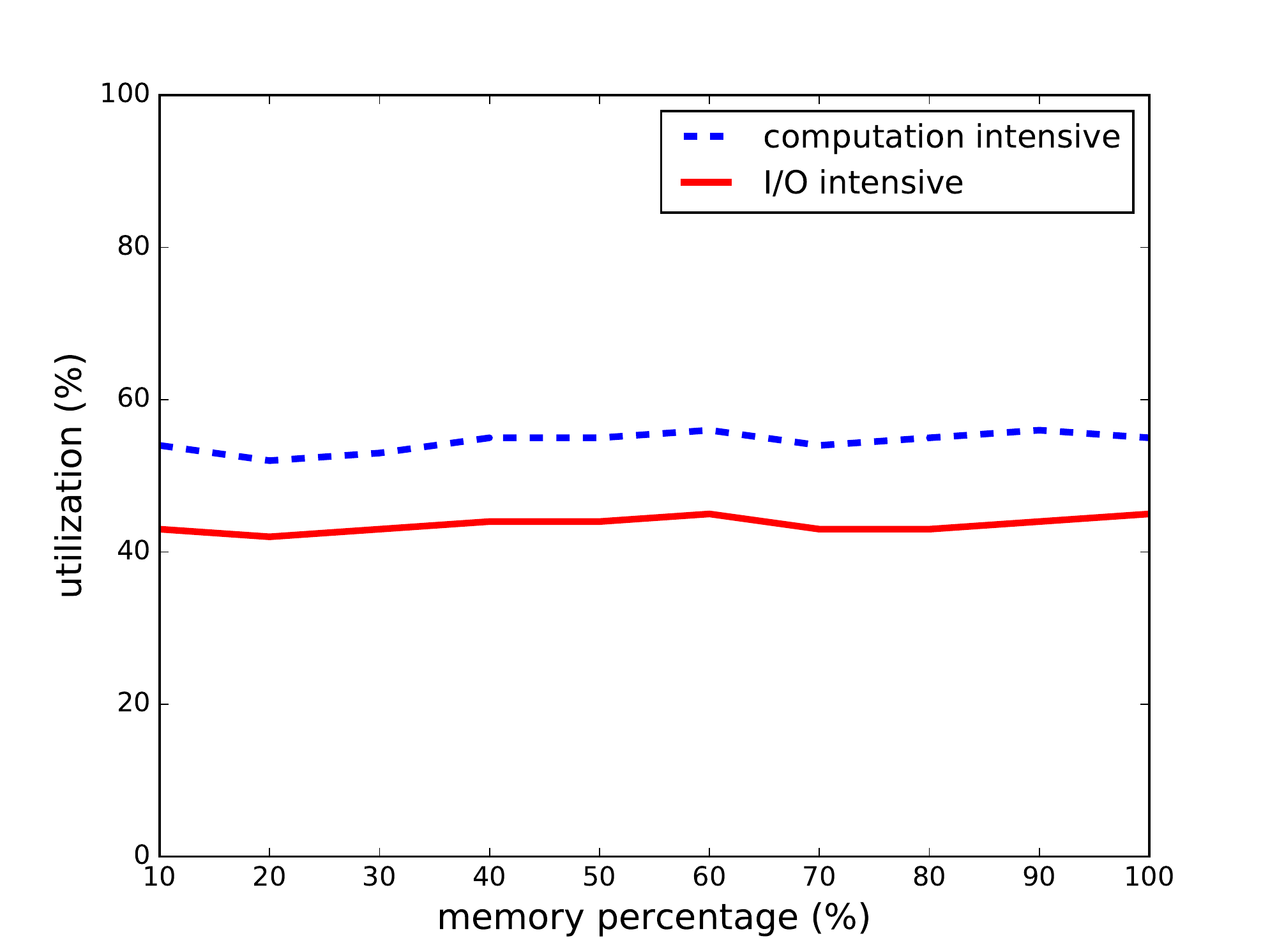}}
  \caption{Single execution result}
  \label{fig:subfig} %% label for entire figure
\end{figure}

We test both computation-intensive and IO-intensive applications with a limited GPU memory percentage, and we measure the how the allocated memory size affect execution time and GPU utilization. The results are surprising, as shown in Figure 7. Allocating too little GPU memory to applications leads to 'out of memory' error, so there is no time and utilization with little memory percentage in Figure 7. The execution time and GPU utilization of both applications stay almost the same, regardless of how much GPU memory is used. The results imply that once the application get enough memory to finish the execution, the performance cannot be improved with more GPU memory.

We then execute two applications concurrently on one GPU and find if the two applications will influence each other. To control variables, we test the computation-intensive application with different memory size when data-intensive application is executed with a fixed memory size, and vice versa. We test the  applications with 25\% and 50\% memory occupied. So the corresponding memory percentage left is up to 75\% and 50\%, as shown in Figure 8. Comparing to running on the GPU singly, the execution time of both applications increases, which proves that the executing several applications on one GPU will cause performance loss. Noted that the execution time of one application still stay the same, regardless of the memory size allocated to the other application. Even though there is available GPU memory space, the applications may sustain performance loss. Through the results shown in Figure 7 and Figure 8, we draw the conclusion that although Tensorflow applications may occupy all GPU memory if not limited, the memory size actually has little influence on execution performance.

\begin{figure}[tbp]
  \centering
  \subfigure[Result of IO-intensive application]{
    \label{fig:subfig:a} %% label for first subfigure
    \includegraphics[width=0.40\textwidth]{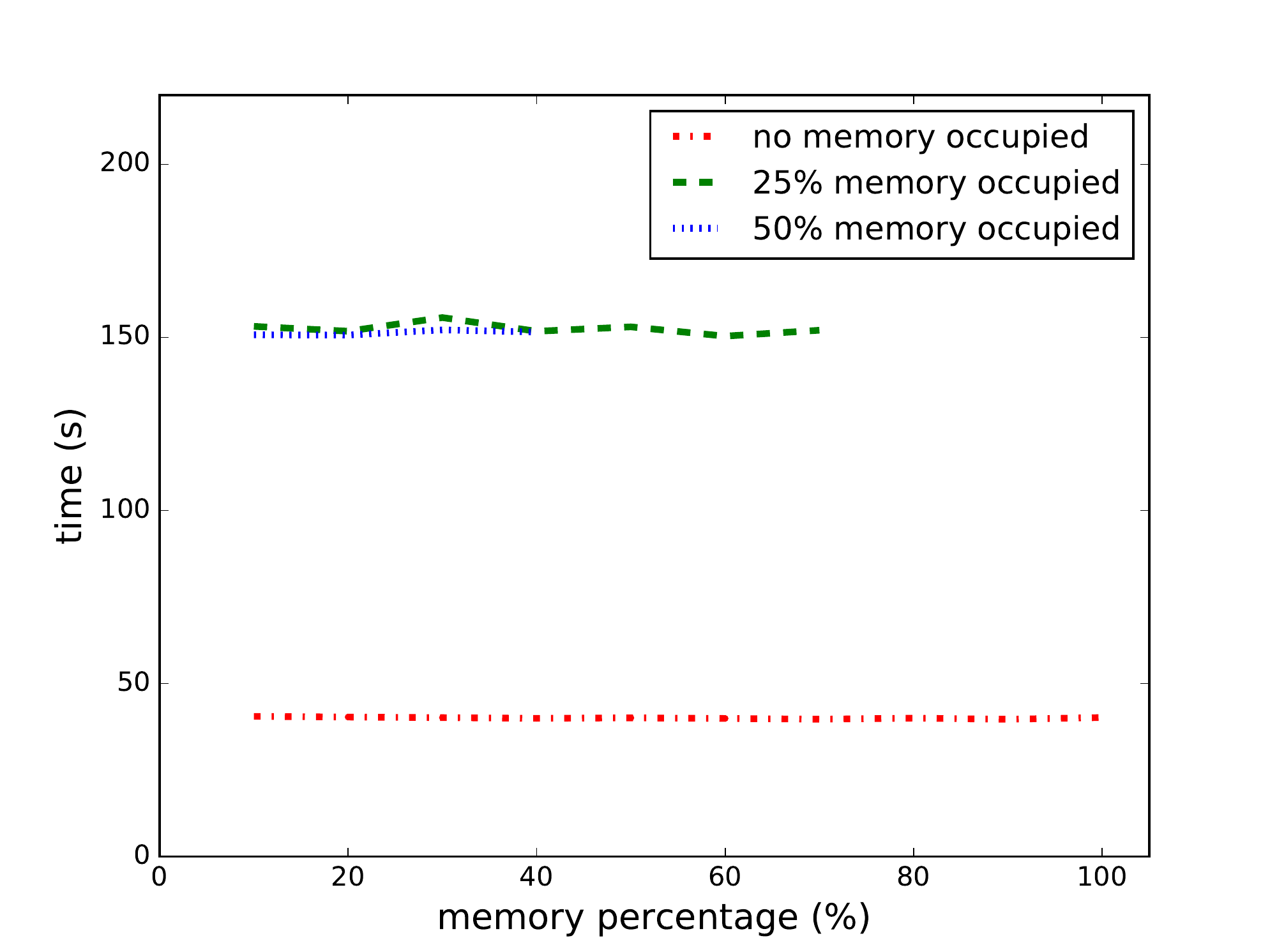}}
  \hspace{0.01in}
  \subfigure[Result of computation-intensive application]{
    \label{fig:subfig:b} %% label for second subfigure
    \includegraphics[width=0.40\textwidth]{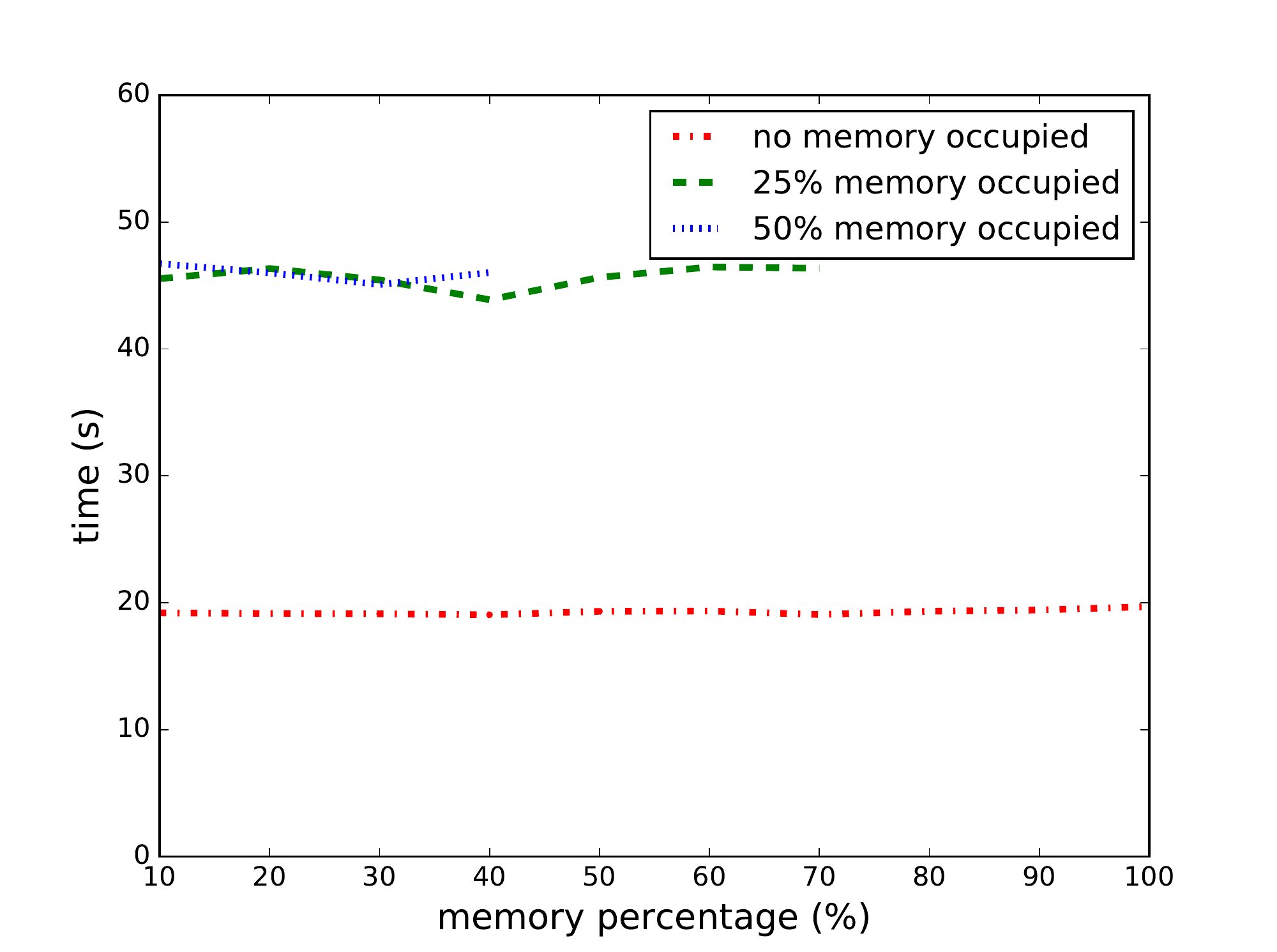}}
  \caption{Concurrent execution result}
  \label{fig:subfig} %% label for entire figure
\end{figure}
Based the above analysis, we find that although Tensorflow applications intend to use as much GPU as they can, high memory usage does not imply high performance.

\subsection{Performance Diagnosis}
\texttt{DeepProf} can also be used for simple performance diagnosis. We show two typical performance issues that can be detected using \texttt{DeepProf}.

The first case is the graph growth issues. As mentioned in Section \uppercase\expandafter{\romannumeral3}, adding new nodes to the dataflow graph leads to graph re-initialization. In fact, the session has to initialize the graph before execution. Without graph change, the session only needs to initialize the graph once in the first iteration and keeps the initialized graph for later invocation. The graph growth causes the session to initialize the graph in every iteration and produce a huge overhead. \texttt{DeepProf} can help developers to diagnose such performance issues. Figure 9 (a) shows the execution state graph of the example program in Figure 3, along with the metrics calculated by \texttt{DeepProf}. We can find that the \textit{avg interval} is much larger than \textit{avg operation}, implying the performance bottleneck exists outside the iteration. In other words, no GPU computing operation somehow blocks the underlying execution. As the state graph shows, both slow data copy and CPU instructions may cause such block. In Figure 9 (a), the \textit{avg overlap} metric, representing the average ratio of data copy time during the iteration interval, is small. Small \textit{avg overlap} implies the block is not caused by data copy operations since data copy only contributes little to the interval time. Therefore, it is very likely that the CPU instructions are the origin of the issue and developers can check the training loop part of source codes. \texttt{DeepProf} successfully detects the graph growth issues and help locating the slow code in this case.
\begin{figure}[tbp]
  \centering
  \subfigure[Case 1 results]{
    \label{fig:subfig:a} %% label for first subfigure
    \includegraphics[width=0.23\textwidth]{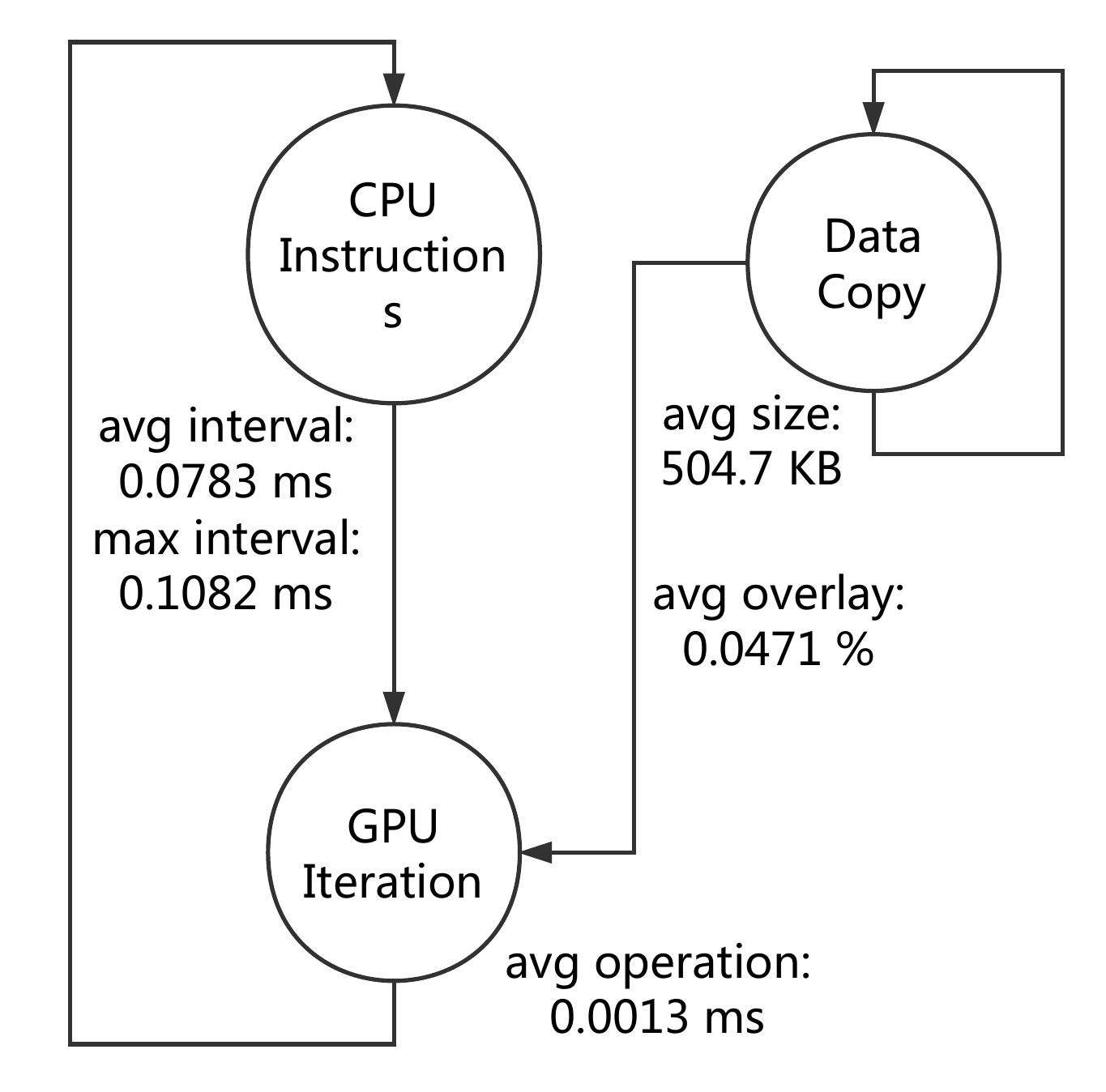}}
  \hspace{0.01in}
  \subfigure[Case 2 results]{
    \label{fig:subfig:b} %% label for second subfigure
    \includegraphics[width=0.23\textwidth]{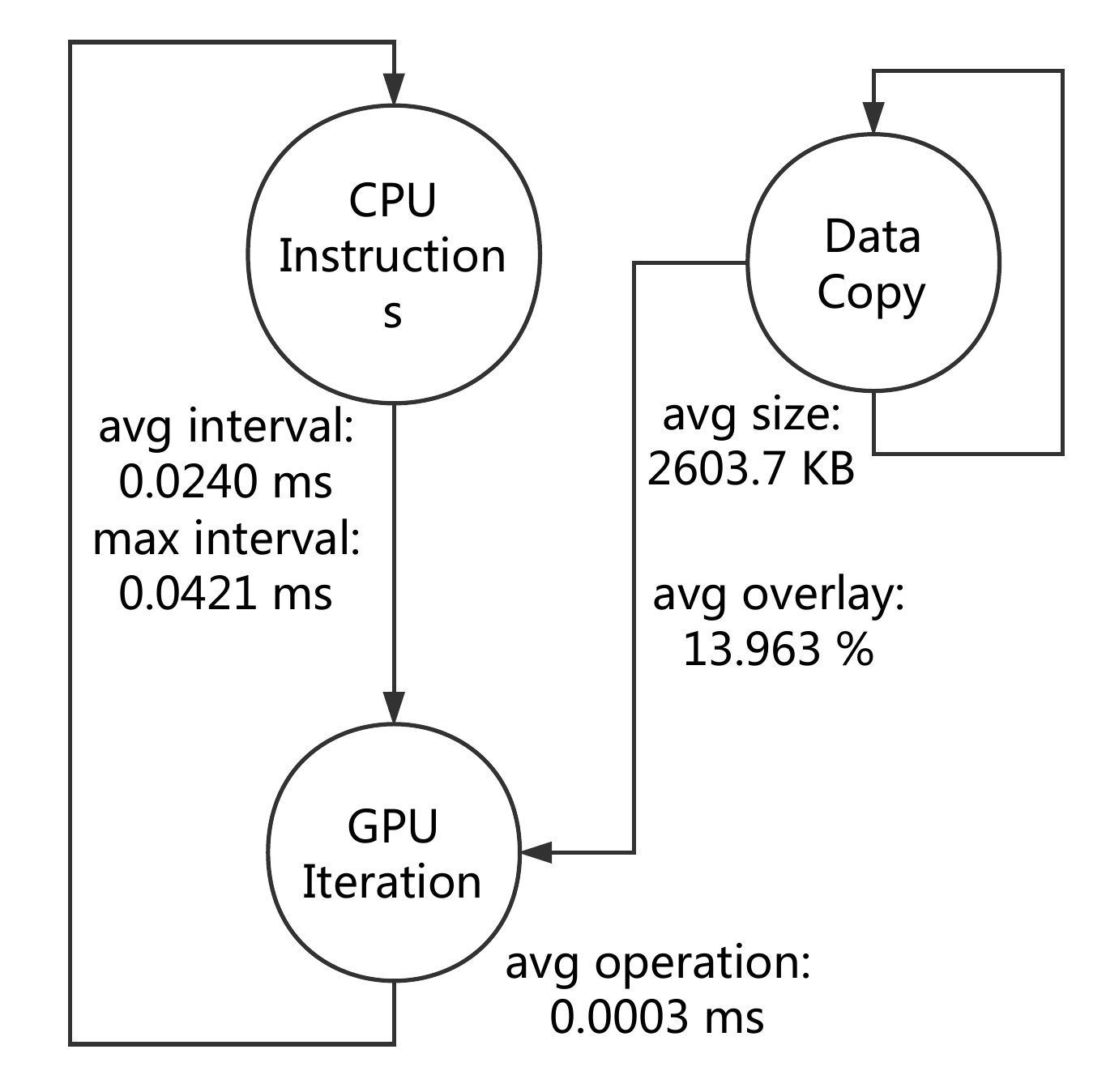}}
  \caption{Results of \texttt{DeepProf} for performance diagnosis}
  \label{fig:subfig} %% label for entire figure
\end{figure}

The second case is the oversize data copy. Data transfer is a well-known bottleneck in deep learning field. Deep learning applications are generally designed to process tremendous data but the data transfer efficiency between host and GPU can't match the great computing power of GPU. Although data transfer efficiency is an inherent problem, performance issues can be unexpectedly caused by inappropriate parameters. Deep learning applications usually process the data in batch, thus improper batch size may cause unnecessary overheads.
Figure 9 (b) shows the execution state graph of an application with an overlarge batch size. In the state graph, we can see \textit{avg interval} is large, and so does \textit{avg overlap}. This phenomenon implies that the GPU iteration is blocked by data copy operations, since 12 percent of interval time is spent on waiting data copy operations. A smaller batch size in the application can improve the execution performance. On the other hand, some deep learning algorithms require certain amount of data for processing and the batch size cannot be smaller. In this case, developers should decide on the tradeoff between performance and effectiveness. \texttt{DeepProf} diagnose the data copy bottleneck in this case.

The two cases above are all based on real Tensorflow applications and inexperienced developers often make such mistakes. \texttt{DeepProf} can help these inexperienced developers to rapidly diagnose the application performance.
\subsection{Discussions}
Through the empirical study, \texttt{DeepProf} shows the power of analyzing the performance of deep learning applications. \texttt{DeepProf} is also capable of detecting the common performance bottleneck and help developers to identify mistakes. The space complexity of \texttt{DeepProf} is $O(n)$, which means \texttt{DeepProf} can deal with large GPU traces. Moreover, the results generated by \texttt{DeepProf} are based on the number of iterations of the application and GPU traces collected by \textit{nvprof}, thus the design ideas of \texttt{DeepProf} can be applied to most GPU-based deep learning frameworks.

However, there are also some defects in \texttt{DeepProf}. The performance metrics generated by \texttt{DeepProf} are straightforward and how to find more subtle metrics with the partitioned GPU trace is worth studying. In addition, \texttt{DeepProf} focuses on analyzing the performance of applications executed on single GPU device. Analyzing the performance of deep learning applications under multi-GPU circumstance remains an open question.

\section{related work}

\subsection{Profiling}
The usefulness of profiling has long been recognized. Profiling is a kind of dynamic program analysis which is widely used in functional fault detection \cite{bug6, bug7, profile1}, and non-functional fault detection \cite{profile5,profile6,bug3, performance3}. Jiang \emph{et al.} utilized execution profilers that possibly contain faults to simplify programs and scale down the complexity of programs for in-house testing \cite{profile7}. AppInsight, provided by Ravindranath \emph{et al.}, instruments mobile-app binaries to identify the critical path in user transactions automatically \cite{profile4}. Coppa \emph{et al.} proposed an approach to measure how the size of input effects performance, and used it to find out performance faults by analyzing profiles \cite{nfd_profile1}. Chilimbi \emph{et al.} provided HOLMES, a tool to profile the selected parts of the application, and then rank and identify the critical paths which can predict the failures through path profiles \cite{profile1}. Han \emph{et al.} proposed StackMine, which extracts effective subsequences of function calls by a costly-pattern mining algorithm, to help the performance debugging. Shen \emph{et al.} proposed GA-Prof, which used search-based input-sensitive application profiling for automating performance bottlenecks detection \cite{performance_profile}. These work identify critical paths and help identify performance problems through profiling. \texttt{DeepProf} takes the core idea of profiling analysis and focus on handling GPU traces.

Several work focus on deriving operational profiles with clustering algorithms \cite{operational1}. Gittens \emph{et al.} increased the profiling applicability by an extended operational profile model to address the heterogeneity of software. Nagappan \emph{et al.} proposed a suffix-array based algorithm to parse the execution logs and generate operational profiles \cite{profile3}. However, GPU traces are much more complicated than logs since one upper API call may cause several GPU operations. Furthermore, \texttt{DeepProf} provides execution summaries from GPU traces which can help developers to analyze application performance.

\subsection{Detecting and fixing performance problems}
Detecting and fixing performance problems were shown to be challenging \cite{performance0}. Several techniques have been proposed to identify performance problems such as slow code \cite{bug3, performance16, traceperformance, bug1, bug2} and increasing execution time \cite{nfd_profile1, performance_profile, performance12}. Grechanik \emph{et al.} proposed FOREPOST, a feedback-directed black-box approach for detecting performance problems and identifying bottlenecks \cite{performance4}. Liu \emph{et al.} designed an innovative system, AutoAnalyzer, to identify existence of performance bottlenecks by clustering algorithms and to locate the bottlenecks using searching algorithm \cite{performance9}. Song and Lu investigated design points in statistical debugging for problem diagnosis \cite{bug}. Chis pinpointed memory issues through detecting memory anti-patterns from memory catalogs \cite{performance_pattern2}. Nistor \emph{et al.} designed Toddler, which detecting performance bugs through similar memory-access patterns \cite{performance_pattern3}. Chen \emph{et al.} proposed a framework to detect object-relational mapping performance anti-patterns automatically \cite{performance_pattern1}. Other work focused on concurrency performance problems \cite{bug5}, performance testing \cite{performance9} and latent performance bugs \cite{performance8}. \texttt{DeepProf} is designed for analyzing performance of deep learning applications through mining patterns in GPU traces, and such scenario is not covered by previous work. To the best of our knowledge, we are the first to analyze the performance of GPU-based deep learning applications.

\section{Conclusion}
Deep learning applications are computation-intensive in nature and may incur incredibly lengthy execution time. Although powerful frameworks, like Tensorflow, make it convenient to develop deep learning applications, the complex underlying GPU operations make it difficult to measure the performance of applications. In this paper, through analyzing the execution procedure of Tensorflow applications in detail, we propose a novel analysis tool, \texttt{DeepProf}, to automatically mine the patterns from GPU traces and analyze application performance. We further verify the effectiveness of \texttt{DeepProf} through the empirical study on performance analysis and diagnosis. We also conclude the underlying execution features of Tensorflow, which can be used as guidance for the deep learning system setup. To the best of out knowledge, we are the first to analyze the performance of GPU-based deep learning applications. Although \texttt{DeepProf} is designed for Tensorflow applications, the preprocessing procedure of GPU traces can adapt to all GPU-based deep learning applications. Finally, in our future work, we are interested in extending \texttt{DeepProf} to analyze the performance of application using multi-GPU and to achieve bug detection by taking CPU profiling into consideration.


%\section*{Acknowledgment}




%\bibliographystyle{IEEEtran}
{
\bibliographystyle{IEEEtran}
\bibliography{performance}

\begin{thebibliography}{10}
\providecommand{\url}[1]{#1}
\csname url@samestyle\endcsname
\providecommand{\newblock}{\relax}
\providecommand{\bibinfo}[2]{#2}
\providecommand{\BIBentrySTDinterwordspacing}{\spaceskip=0pt\relax}
\providecommand{\BIBentryALTinterwordstretchfactor}{4}
\providecommand{\BIBentryALTinterwordspacing}{\spaceskip=\fontdimen2\font plus
\BIBentryALTinterwordstretchfactor\fontdimen3\font minus
  \fontdimen4\font\relax}
\providecommand{\BIBforeignlanguage}[2]{{%
\expandafter\ifx\csname l@#1\endcsname\relax
\typeout{** WARNING: IEEEtran.bst: No hyphenation pattern has been}%
\typeout{** loaded for the language `#1'. Using the pattern for}%
\typeout{** the default language instead.}%
\else
\language=\csname l@#1\endcsname
\fi
#2}}
\providecommand{\BIBdecl}{\relax}
\BIBdecl

\bibitem{deep}
Y.~LeCun, Y.~Bengio, and G.~Hinton, ``Deep learning,'' \emph{Nature}, vol. 521,
  no. 7553, pp. 436--444, 2015.

\bibitem{audio}
G.~Hinton, L.~Deng, D.~Yu, G.~E. Dahl, A.-r. Mohamed, N.~Jaitly, A.~Senior,
  V.~Vanhoucke, P.~Nguyen, T.~N. Sainath \emph{et~al.}, ``Deep neural networks
  for acoustic modeling in speech recognition: The shared views of four
  research groups,'' \emph{IEEE Signal Processing Magazine}, vol.~29, no.~6,
  pp. 82--97, 2012.

\bibitem{nlp}
R.~Collobert and J.~Weston, ``A unified architecture for natural language
  processing: Deep neural networks with multitask learning,'' in \emph{Proc. of
  the 25th ACM International Conference on Machine Learning (ICML).}, 2008, pp.
  160--167.

\bibitem{cv}
A.~Krizhevsky, I.~Sutskever, and G.~E. Hinton, ``Imagenet classification with
  deep convolutional neural networks,'' in \emph{Advances in neural information
  processing systems}, 2012, pp. 1097--1105.

\bibitem{tf}
M.~Abadi, P.~Barham, J.~Chen, Z.~Chen, A.~Davis, J.~Dean, M.~Devin,
  S.~Ghemawat, G.~Irving, M.~Isard \emph{et~al.}, ``Tensorflow: A system for
  large-scale machine learning,'' in \emph{Proc. of the 12th USENIX Symposium
  on Operating Systems Design and Implementation (OSDI).}, 2016.

\bibitem{nvidia}
``Nvidia gpus,'' \url{http://developer.nvidia.com/deep-learning/}.

\bibitem{tfurl}
``Tensorflow,'' \url{https://www.tensorflow.org/}.

\bibitem{nvprofiler}
``Gpu profiler,''
  \url{http://docs.nvidia.com/cuda/profiler-users-guide/index.html/}.

\bibitem{weiner1973linear}
P.~Weiner, ``Linear pattern matching algorithms,'' in \emph{Proc. of the IEEE
  Conference Record of 14th Annual Symposium on Switching and Automata Theory
  (SWAT).}, 1973, pp. 1--11.

\bibitem{alpaydin2014introduction}
E.~Alpaydin, \emph{Introduction to machine learning}.\hskip 1em plus 0.5em
  minus 0.4em\relax MIT press, 2014.

\bibitem{machinelearning}
R.~Kohavi and F.~Provost, ``Glossary of terms,'' \emph{Machine Learning},
  vol.~30, no. 2-3, pp. 271--274, 1998.

\bibitem{pr}
C.~M. Bishop, ``Pattern recognition,'' \emph{Machine Learning}, vol. 128, pp.
  1--58, 2006.

\bibitem{theano}
``Theano,'' \url{http://deeplearning.net/software/theano/index.html}.

\bibitem{cuda}
``Cuda toolkit,'' \url{https://developer.nvidia.com/cuda-toolkit/}.

\bibitem{nvprof}
``nvprof tool,'' \url{http://docs.nvidia.com/cuda/profiler-users-guide}.

\bibitem{bug6}
M.~Pradel and T.~R. Gross, ``Leveraging test generation and specification
  mining for automated bug detection without false positives,'' in \emph{Proc.
  of the 34th IEEE International Conference on Software Engineering (ICSE).},
  2012, pp. 288--298.

\bibitem{bug7}
S.~Zhang and M.~D. Ernst, ``Automated diagnosis of software configuration
  errors,'' in \emph{Proc. of the 35th IEEE International Conference on
  Software Engineering (ICSE).}, 2013, pp. 312--321.

\bibitem{profile1}
T.~M. Chilimbi, B.~Liblit, K.~Mehra, A.~V. Nori, and K.~Vaswani, ``Holmes:
  Effective statistical debugging via efficient path profiling,'' in
  \emph{Proc. of the 31st IEEE International Conference on Software Engineering
  (ICSE).}, 2009, pp. 34--44.

\bibitem{profile5}
D.~Yan, G.~Xu, and A.~Rountev, ``Uncovering performance problems in java
  applications with reference propagation profiling,'' in \emph{Proc. of the
  34th IEEE International Conference on Software Engineering (ICSE).}, 2012,
  pp. 134--144.

\bibitem{profile6}
G.~Xu and A.~Rountev, ``Precise memory leak detection for java software using
  container profiling,'' in \emph{Proc. of the 30th ACM/IEEE International
  Conference On Software Engineering (ICSE).}, 2008, pp. 151--160.

\bibitem{bug3}
S.~Han, Y.~Dang, S.~Ge, D.~Zhang, and T.~Xie, ``Performance debugging in the
  large via mining millions of stack traces,'' in \emph{Proc. of the 34th IEEE
  International Conference on Software Engineering (ICSE).}, 2012, pp.
  145--155.

\bibitem{performance3}
B.~Chen, Y.~Liu, and W.~Le, ``Generating performance distributions via
  probabilistic symbolic execution,'' in \emph{Proceedings of the 38th
  International Conference on Software Engineering}.\hskip 1em plus 0.5em minus
  0.4em\relax ACM, 2016, pp. 49--60.

\bibitem{profile7}
L.~Jiang and Z.~Su, ``Profile-guided program simplification for effective
  testing and analysis,'' in \emph{Proc. of the 16th ACM International
  Symposium on Foundations of software engineering (SIGSOFT).}, 2008, pp.
  48--58.

\bibitem{profile4}
L.~Ravindranath, J.~Padhye, S.~Agarwal, R.~Mahajan, I.~Obermiller, and
  S.~Shayandeh, ``Appinsight: Mobile app performance monitoring in the wild.''
  in \emph{Proc. of the 12th USENIX Symposium on Operating Systems Design and
  Implementation (OSDI).}, vol.~12, 2012, pp. 107--120.

\bibitem{nfd_profile1}
E.~Coppa, C.~Demetrescu, and I.~Finocchi, ``Input-sensitive profiling,''
  \emph{ACM SIGPLAN Notices}, vol.~47, no.~6, pp. 89--98, 2012.

\bibitem{performance_profile}
D.~Shen, Q.~Luo, D.~Poshyvanyk, and M.~Grechanik, ``Automating performance
  bottleneck detection using search-based application profiling,'' in
  \emph{Proc. of the ACM International Symposium on Software Testing and
  Analysis (ISSTA)}, 2015, pp. 270--281.

\bibitem{operational1}
J.~D. Musa, ``Operational profiles in software-reliability engineering,''
  \emph{IEEE software}, vol.~10, no.~2, pp. 14--32, 1993.

\bibitem{profile3}
M.~Nagappan, K.~Wu, and M.~A. Vouk, ``Efficiently extracting operational
  profiles from execution logs using suffix arrays,'' in \emph{Proc. of the
  20th IEEE International Symposium on Software Reliability Engineering
  (ISSRE).}, 2009, pp. 41--50.

\bibitem{performance0}
G.~Jin, L.~Song, X.~Shi, J.~Scherpelz, and S.~Lu, ``Understanding and detecting
  real-world performance bugs,'' \emph{ACM SIGPLAN Notices}, vol.~47, no.~6,
  pp. 77--88, 2012.

\bibitem{performance16}
X.~Xiao, S.~Han, D.~Zhang, and T.~Xie, ``Context-sensitive delta inference for
  identifying workload-dependent performance bottlenecks,'' in \emph{Proc. of
  the ACM International Symposium on Software Testing and Analysis (ISSTA).},
  2013, pp. 90--100.

\bibitem{traceperformance}
X.~Yu, S.~Han, D.~Zhang, and T.~Xie, ``Comprehending performance from
  real-world execution traces: A device-driver case,'' in \emph{ACM SIGPLAN
  Notices}, vol.~49, no.~4, 2014, pp. 193--206.

\bibitem{bug1}
C.~Sauvanaud, K.~Lazri, M.~Ka{\^a}niche, and K.~Kanoun, ``Anomaly detection and
  root cause localization in virtual network functions,'' in \emph{Proc. of the
  27th IEEE International Symposium on Software Reliability Engineering
  (ISSRE).}, 2016, pp. 196--206.

\bibitem{bug2}
W.~Wen, T.~Yu, and J.~H. Hayes, ``Colua: Automatically predicting configuration
  bug reports and extracting configuration options,'' in \emph{Proc. of the
  27th IEEE International Symposium on Software Reliability Engineering
  (ISSRE).}, 2016, pp. 150--161.

\bibitem{performance12}
A.~Nistor, P.-C. Chang, C.~Radoi, and S.~Lu, ``Caramel: detecting and fixing
  performance problems that have non-intrusive fixes,'' in \emph{Proc. of the
  37th IEEE International Conference on Software Engineering (ICSE).}, 2015,
  pp. 902--912.

\bibitem{performance4}
M.~Grechanik, C.~Fu, and Q.~Xie, ``Automatically finding performance problems
  with feedback-directed learning software testing,'' in \emph{Proc. of the
  34th IEEE International Conference on Software Engineering (ICSE).}, 2012,
  pp. 156--166.

\bibitem{performance9}
X.~Liu, J.~Zhan, K.~Zhan, W.~Shi, L.~Yuan, D.~Meng, and L.~Wang, ``Automatic
  performance debugging of spmd-style parallel programs,'' \emph{Journal of
  Parallel and Distributed Computing}, vol.~71, no.~7, pp. 925--937, 2011.

\bibitem{bug}
L.~Song and S.~Lu, ``Statistical debugging for real-world performance
  problems,'' \emph{ACM SIGPLAN Notices}, vol.~49, no.~10, pp. 561--578, 2014.

\bibitem{performance_pattern2}
A.~E. Chis, ``Automatic detection of memory anti-patterns,'' in \emph{Companion
  to the 23rd ACM SIGPLAN conference on Object-oriented programming systems
  languages and applications}, 2008, pp. 925--926.

\bibitem{performance_pattern3}
A.~Nistor, L.~Song, D.~Marinov, and S.~Lu, ``Toddler: Detecting performance
  problems via similar memory-access patterns,'' in \emph{Proc. of the 35th
  IEEE International Conference on Software Engineering (ICSE).}, 2013, pp.
  562--571.

\bibitem{performance_pattern1}
T.-H. Chen, W.~Shang, Z.~M. Jiang, A.~E. Hassan, M.~Nasser, and P.~Flora,
  ``Detecting performance anti-patterns for applications developed using
  object-relational mapping,'' in \emph{Proc. of the 36th ACM International
  Conference on Software Engineering (ICSE).}, 2014, pp. 1001--1012.

\bibitem{bug5}
J.~Oh, C.~J. Hughes, G.~Venkataramani, and M.~Prvulovic, ``Lime: A framework
  for debugging load imbalance in multi-threaded execution,'' in \emph{Proc. of
  the 33rd ACM International Conference on Software Engineering (ICSE).}, 2011,
  pp. 201--210.

\bibitem{performance8}
C.~Killian, K.~Nagaraj, S.~Pervez, R.~Braud, J.~W. Anderson, and R.~Jhala,
  ``Finding latent performance bugs in systems implementations,'' in
  \emph{Proc. of the 18th ACM International Symposium on Foundations of
  Software Engineering (SIGSOFT).}, 2010, pp. 17--26.

\end{thebibliography}
}
%\begin{thebibliography}{1}


%\end{thebibliography}

\end{document}